# Solution based synthesis of mixed-phase materials in the $Li_2TiO_3$ - $Li_4SiO_4$ system.

*Dorian A. H. Hanaor[(1)\*], Matthias H. H. Kolb[(2)], Yixiang Gan[(1)], Marc Kamlah[(2)], Regina Knitter[(2)]*

(1)     School of Civil Engineering, University of Sydney, NSW 2006, Australia

(2)     Institute for Applied Materials, Karlsruhe Institute of Technology, 76021 Germany

**Abstract:**

As candidate tritium breeder materials for use in the ITER helium cooled pebble bed, ceramic multiphasic compounds lying in the region of the quasi-binary lithium metatitanate- lithium orthosilicate system may exhibit mechanical and physical advantages relative to single phase materials. Here we present an organometallic solution-based synthesis procedure for the low-temperature fabrication of compounds in the $Li_2TiO_3$-$Li_4SiO_4$ region and investigate phase stability and transformations through temperature varied X-ray diffraction and scanning calorimetry. Results demonstrate that the metatitanate and metasilicate phases $Li_2TiO_3$ and $Li_2SiO_3$ readily crystallise in nanocrystalline form at temperatures below 180°C. Lithium deficiency in the region of 5% results from Li sublimation from $Li_4SiO_4$ and/or from excess Li incorporation in the metatitanate phase and brings about a stoichiometry shift and product compounds with mixed lithium orthosilicate/ metasilicate content towards the Si rich region and predominantly $Li_2TiO_3$ content towards the Ti rich region. Above 1150°C the transformation of monoclinic to cubic γ-$Li_2TiO$ disordered solid-solution occurs while the melting of silicate phases indicates a likely monotectic type system with a solidus line in the region 1050°-1100°C. Synthesis procedures involving a lithium chloride precursor are not likely to be a viable option for breeder pebble synthesis as this route was found to yield materials with a more significant Li-deficiency exhibiting the crystallisation of the $Li_2TiSiO_5$ phase at intermediate compositions.

*Corresponding author: email dorian.hanaor@sydney.edu.au, ph. +61-404-188810

## *1.* *Introduction:*

A key component of the ITER thermonuclear demonstrator reactor is the helium cooled pebble bed (HCPB) breeding blanket that utilises Li bearing ceramic pebbles for tritium breeding purposes[1-3]. Numerous candidate oxide compounds have been studied towards this goal. Material requirements include adequate thermo-mechanical stability, low activation levels, and sufficiently high lithium density. In recent years lithium metatitanate ($Li_2TiO_3$) and lithium orthosilicate ($Li_4SiO_4$), in some cases with Li excess[4-7], have emerged as leading HCPB candidate materials, as these compounds are found to exhibit low activation relative to earlier studied compositions of $Li_2ZrO_3$ and $LiAlO_2$ while maintaining favourable thermo-mechanical properties [1, 8-10]. The properties exhibited by candidate pebble bed materials are the subject of intensive investigation and review as these are critical to the viability and performance of the numerous solid breeder conceptual designs [11]. Consequently increased versatility in the attainable properties of pebble material can be expected to yield meaningful benefits for breeder blanket design.

Research of breeder pebbles produced using melt-based processing methods involving the addition of $TiO_2$ to $Li_4SiO_4$ melt has shown that materials comprised of biphasic mixtures of $Li_4SiO_4$ and $Li_2TiO_3$ exhibit improved mechanical properties relative to single phase materials, evident by increased crush stress [12, 13]. However, the stoichiometric and thermal limitations imposed by melt-processing restrict the range of compositions and microstructures attainable in the Li rich region of the ternary $Li_2O$-$SiO_2$-$TiO_2$ system. For this reason the wet-chemical processes utilising precursor compounds in the form of salts and/or organometallic compounds to yield homogenous biphasic or triphasic oxide materials presents an attractive approach to breeder materials syntheses.

A number of recent studies have investigated methods for the fabrication of single phase $Li_2TiO_3$ and $Li_4SiO_4$ pebbles from powder-form precursor materials in dry and wet processes utilising binders and different spheroidisation techniques [14-20]. Such methods readily allow the fabrication of pebbles with well controlled geometry, density and consistency and reprocessing by melting and by wet chemical methods has further been the subject of various research efforts [21, 22]. In parallel, multiple solution based syntheses for lithium metatitanate and orthosilicate have been reported to date [23-28]. Such processes generally utilise organometallic compounds and/or salts as Si, Ti and Li precursors for the synthesis of oxides of various stoichiometry. Relative to solid state processing, such syntheses have the advantage of better mixing and consequently shorter diffusion distances, imparting greater flexibility with respect to thermal processing, potentially allowing lower temperature synthesis of the desired oxide compounds, attainment of finer crystallite sizes and compositional versatility.

$Li_2TiO_3$ is a congruently melting phase exhibiting high temperature polymorphism while $Li_4SiO_4$ exists in a monoclinic structure up to temperatures of 1258°C at which this phase melts congruently, although earlier studies reported a peritectic decomposition at 1255°C [29-32]. The metasilicate phase is also studied for potential applications in breeder pebbles and melts congruently at 1209°C [30, 33]. As both endpoints are congruently melting phases, the $Li_2TiO_3$-$Li_4SiO_4$ composition range is considered to be a quasi-binary system. A summary of phase equilibria reported in the literature for the $Li_2O$-$TiO_2$-$SiO_2$ (LTS) ternary system is shown in Figure 1 [30-32, 34-37]. In this figure the liquidus temperatures are taken from the most recent available references. Compositions 1-11 studied in the present work are further indicated in the diagram. Understanding of stability and equilibria in this system remains to date incomplete in the Li-rich ternary regime as tie-lines, liquidus surfaces and potential eutectic/eutectoid compositions have not been determined comprehensively in this region of the system. Reported results suggest a likely compatibility tie line between $Li_4SiO_4$ and $Li_2TiO_3$ as these two phases are found to coexist in melt-based products [12]. A further compatibility tie line or potentially mutual solubility is likely to exist between the ortho-phases (66.67 mol% $Li_2O$) in the system owing to these compounds exhibiting similar cation-oxygen bond lengths and the isostructural relationship between high temperature β-$Li_4TiO_4$ (stable above 686 °C) and $Li_4SiO_4$ [31, 38-40].

Consequently, in an effort to investigate the synthesis of breeder materials of new compositions, we seek to explore the wet-chemical synthesis of compounds in what is likely to be a binary $Li_2TiO_3$-$Li_4SiO_4$ system.

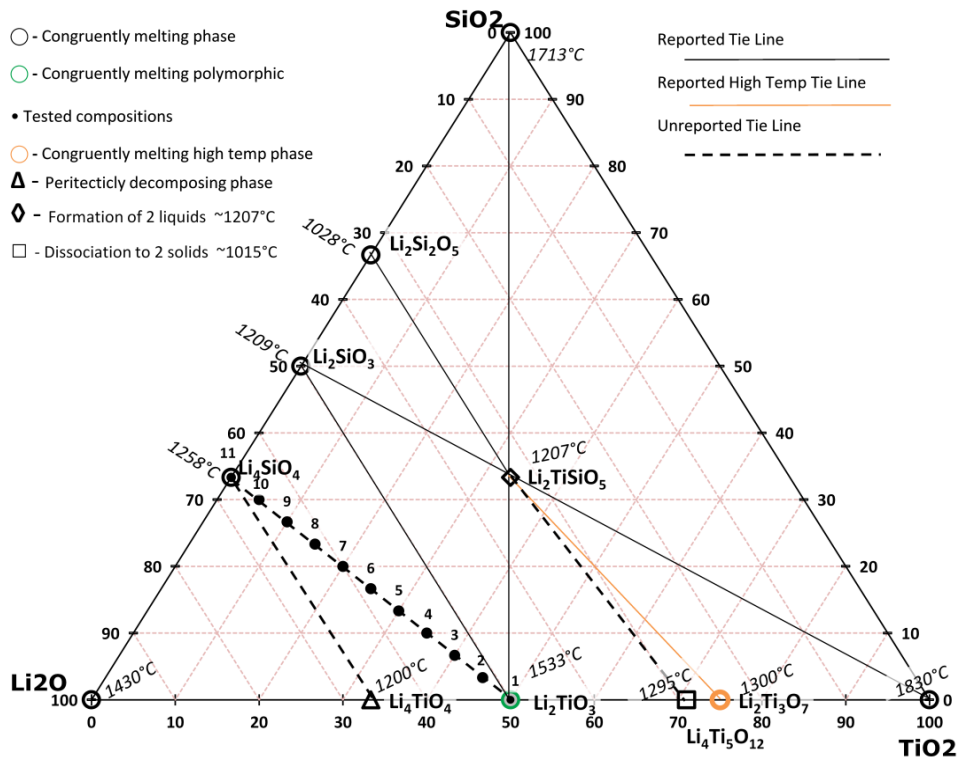

**Figure 1. Composite diagram of reported phase equilibria in the Li₂O-TiO₂-SiO₂ system [30-32, 34-37], indicating liquidus temperatures and the positions of compositions 1-11 targeted in the present work.**

## 2. Materials and Methods:

All sols fabricated in the presently reported work were ethanol based. LiOH monohydrate and LiCl (both Alfa Aesar, 99%) were examined for use as Li precursor compounds following similar methods for solution processing of metal oxides [23, 27]. Prior to use, LiCl and LiOH·H₂O were vacuum dried at 180°C yielding anhydrous products as confirmed by mass balance. Organometallic Ti and Si precursor compounds used were respectively titanium tetra-isopropoxide (TTIP, 97%, Sigma Aldrich), tetraethyl orthosilicate (TEOS, 99%, Sigma Aldrich).

11 Compositions were studied in the present work and these have been denoted according to their equivalent integer ratios of binary oxides as outlined in Table 1. Precursor sols corresponding to the Li₂TiO₃ composition (1:1 Li₂O:TiO₂) are referred to as LT, while sols corresponding to the Li₄SiO₄ composition (2:1 Li₂O:SiO₂) are referred to as L2S. These precursor sols were prepared in separate flasks in quantities totalling 300 g with each with a total oxide content of 0.3 mol. Hence L2S sols contained Li corresponding to 0.2 mol Li₂O, and Si corresponding to 0.1 mol of SiO₂ and LT sols contained Ti and Li precursors equivalent to 0.15 mol each of TiO₂ and Li₂O.

**Table 1: Compositions of materials fabricated**

| Sample | Target Ratio Li₂TiO₃:Li₄SiO₄ | Ti:Si | Primary oxide composition mol / wt % | | |
|---|---|---|---|---|---|
| | | | Li₂O | TiO₂ | SiO₂ |
| 1 | 1 : 0 | 1:0 | 50.00 / 27.2 | 50 / 72.8 | 0 / 0 |
| 2 | 9 : 1 | 27:2 | 51.67 / 28.9 | 45 / 67.3 | 3.33 / 3.8 |
| 3 | 8 : 2 | 6:1 | 53.33 / 30.7 | 40 / 61.6 | 6.67 / 7.7 |
| 4 | 7 : 3 | 21:8 | 55.00 / 32.6 | 35 / 55.5 | 10.00 / 11.9 |
| 5 | 6 : 4 | 9:4 | 56.67 / 34.6 | 30 / 49.0 | 13.33 / 16.4 |
| 6 | 1 : 1 | 3:2 | 58.33 / 36.8 | 25 / 42.1 | 16.67 / 21.1 |
| 7 | 2 : 3 | 1:1 | 60.00 / 39.0 | 20 / 34.8 | 20.00 / 26.2 |
| 8 | 3 : 7 | 9:14 | 61.67 / 41.5 | 15 / 27.0 | 23.33 / 31.5 |
| 9 | 1 : 4 | 3:8 | 63.33 / 44.1 | 10 / 18.6 | 26.67 / 37.3 |
| 10 | 1 : 9 | 1:6 | 65.00 / 46.9 | 5 / 9.6 | 30.00 / 43.5 |
| 11 | 0 : 1 | 0:1 | 66.67 / 49.9 | 0 / 0 | 33.33 / 50.1 |

Chloride LT and L2S sols (those prepared using LiCl as a lithium precursor) were prepared by the gradual addition of LiCl, which had been pre-dissolved in ethanol, to clear ethanolic solutions of TTIP and TEOS under rapid stirring at room temperature. Subsequent to continued stirring for 4 hours, the respective sols were combined at appropriate ratios for compositions 2-10 (compositions 1 and 11 correspond respectively to the original LT and L2S precursor sols). While under stirring, HCl 1M was added to give an alkoxide hydrolysis ratio of R=4 ($H_2O$:(TEOS+TTIP)) to ensure sufficient peptization and avoid the formation of large precipitates. The results were transparent/translucent sols with those containing greater LT content exhibiting higher levels of opacity. Stirring was continued overnight resulting in stable gels which were then dried by evaporation of solvent in ambient air and subsequent heating at 180 °C in a vacuum oven. The resultant xerogel powders were observed to be strongly hygroscopic evident by the rapid appearance of adsorbed aqueous LiCl solution at the materials' surfaces. Subsequent to drying, 500 mg samples of powders were uniaxially pressed at ~293 MPa into 10 mm diameter pellets and calcined at 900°C using a heating rate of 2°/min. The firing temperature of 900°C represents typical operation parameters likely to be encountered in HCPB applications [41].

In subsequent fabrication processes LiOH precursor was employed for the fabrication of colloidal sols corresponding to LT and L2S compositions. This synthesis procedure utilised similar precursor compounds to those employed by Renoult et al. [23] In this synthesis procedure clear 200ml ethanolic solutions of TTIP and TEOS were prepared in separate flasks containing respectively 0.15 moles TTIP and 0.1 moles of TEOS. Hydrolysis of the TTIP solution was achieved by the addition of $HNO_3$ 1M diluted in 50ml of ethanol, under moderate heating, to give a hydrolysis ration of R=2 ($H_2O$:Ti). Acidic pH and heating were used to reduce particle size and prevent the onset of flocculation, dilution of $HNO_3$ in ethanol was undertaken in order to moderate and homogenize the rate of hydrolysis occurring in the solution. Hydrolysis of the ethanolic TEOS solution was achieved similarly using 0.2M $HNO_3$ at R=2. Subsequent to one hour of continued stirring, anhydrous LiOH was gradually added to the LT and L2S sols at respective ratios of Li:Ti=1 and Li:Si=2. Subsequent to approximately 2 hours of continued vigorous stirring, both precursors were in the form of stable colloidal suspensions exhibiting an opaque white appearance. These colloids were left to stir for a further 24 hours after which a distinct orange colouration was evident in both flasks. L2S and LT sols, which were adjusted to 300g by addition of ethanol, were then combined in individual containers following the ratios outlined in table 1. Subsequent to 1 hour of further homogenization by stirring, these were allowed to dry in ambient air, before being dehydrated under vacuum heating at 180°C. In similarity to chloride based materials, dried powders were used to fabricate uniaxially pressed 500 mg pellets which were then fired at 900 °C for 12h with a heating rate of 2°/min. As described in the following section, further analyses of powders fabricated from hydroxide sols were undertaken using temperature varied XRD, TG-DSC and ATR-FTIR.

## 3.      Experimental

The phase composition of fired pelletised samples described above was determined by powder X-ray diffraction, XRD, of crushed samples that was carried out over the range 15-70° 2θ using a Bruker D5000 diffractometer with Cu-Kα emission in ambient air. In order to assess phenomena of crystallisation and phase stability in the synthesised materials fabricated using LiOH precursor, in situ Temperature varied XRD of dried powders was employed using a Bruker D8 Advance diffractometer equipped with an Anton Paar heating chamber. Scans were carried in air out over the range 10-80 °2θ and these were taken at temperature intervals of 50° during heating in the range 100-1200°C and at temperature intervals of 300° during cooling from 1200°C to room temperature. The powders were manually pressed in a platinum-rhodium sample holder to also allow for possible melting of the samples during XRD examination. This facilitated the acquisition of data relating to crystallisation, melting and phase retention in the 11 compositions fabricated. Phase identification was facilitated utilising the ICSD and ICDD PDF-2 databases.

To evaluate the possible formation of organo-lithium products through reaction of lithium hydroxide with solvents or alkoxide precursors, Attenuated Total Reflectance Fourier Transform Infrared spectroscopy (ATR-FTIR) of LT and L2S sols and precipitates was employed using a Bruker Tensor FT-IR spectrometer over the wavenumber range 600-4000 $cm^{-1}$.

To support high temperature XRD analysis, simultaneous thermogravimetry (TG) and differential scanning calorimetry (DSC) of dried powder samples, fabricated using a lithium hydroxide precursor, in the composition range LT-L2S was carried out using a Netzsch STA-449C Jupiter apparatus in ambient air. The reference and the samples were again kept in platinum-alloy crucibles during the experiments. Scans were carried out from room temperature up to 1350 °C with a heating rate of 5° $min^{-1}$ and, subsequent to a dwell time of 1 hour, down to room temperature with a cooling rate of 20° $min^{-1}$.

## 4.    Results
### 4.1.    XRD

XRD spectra from samples fired at 900°C fabricated from chloride and hydroxide precursors for three compositions corresponding to lithium metatitanate (LT), the intermediate composition (L3TS) and lithium orthosilicate (L2S) are shown in Figure 2. It is evident that samples synthesised with the LiCl precursors showed a significant lithium deficiency relative to the target composition. This most likely is the result of the hygroscopic nature of the dried product and the consequent segregation of LiCl to material surfaces and Li loss during heating. Lithium loss through combined lithium rejection at growth interfaces and sublimation of lithium as either $Li_{(g)}$, $LiO_{(g)}$ or $Li_2O_{(g)}$ during firing, particularly above 800°C, is a commonly observed phenomena in Li bearing ceramics [42-48] and thus it is likely to be evident also in materials synthesised from a LiOH precursor, albeit to a notably lesser extent relative to chloride precursor derived samples.

Materials fabricated using lithium chloride exhibited the presence of lithium-poor phases observed by X-ray diffraction. In chloride LT samples the formation of a secondary cubic spinel of composition $Li_{1+x}Ti_{2-x}O_4$, with x=0.33 (frequently referred to as $Li_4Ti_5O_{12}$ [31, 49]), is evident in the (311), (511) and (440) peaks present at 35.6°, 57.2° and 62.8° 2θ respectively [50-52]. This phase was reported by Izquierdo and West in 1980 and by Kleykamp in 2002 as exhibiting a subsolidus decomposition to solid solutions of $Li_2TiO_3$ and $Li_2Ti_3O_7$ at ~1018 °C [31, 36].

The 1:1:1 ternary compound LTS ($Li_2TiSiO_5$), was observed significantly in materials of compositions containing both Si and Ti fabricated using lithium chloride (compositions 2-10). This tetragonal ternary phase was first reported by Kim and Hummel in 1959 and has since been the subject of a small number of further studies [32, 53-55]. The presence of this phase is apparent in the presence of diffraction peaks at 20.2°, 24.5°, 27.7° and 34.5° 2θ corresponding to (001), (101), (200) and (201) planes [53]. At high Ti content, the existence of the LTS phase alongside some titanate spinel may indicate the existence of an as yet unreported compatibility tie-line between these two phases.

Although the synthesis of the tetragonal ternary LTS compound was not the objective of the present work, it should be noted that despite the lower lithium density of this phase (0.22 g $cm^{-3}$ for $LiTiSiO_5$ compared with 0.37 g $cm^{-3}$ for $Li_2TiO_3$, assuming isotopically homogenous $^6Li$), a potentially improved mechanical strength (often observed for congruently melting ternary phases) may confer some advantages with respect to breeder pebble applications.

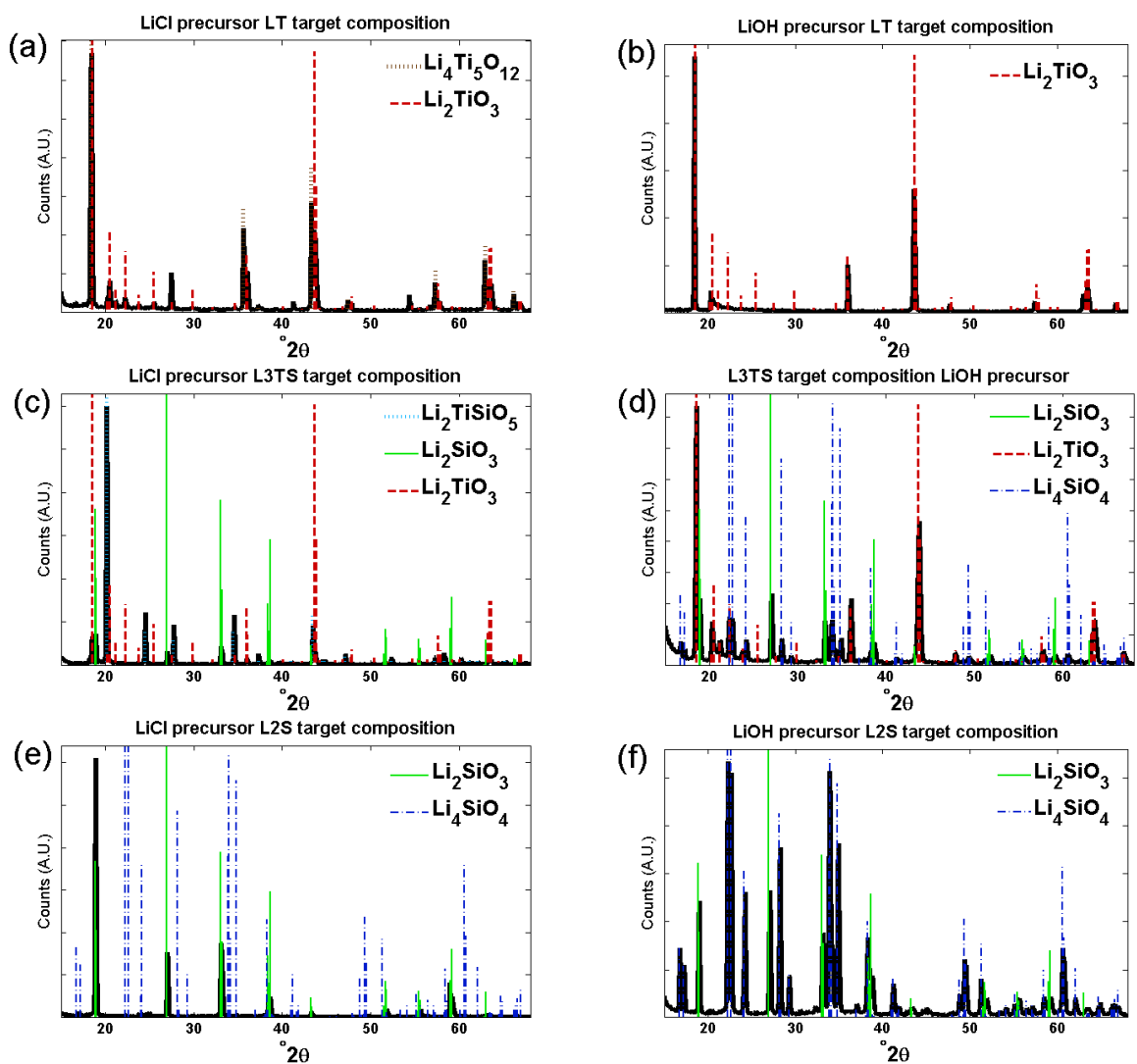

**Figure 2. XRD patterns for pellet samples subsequent to firing at 900°C prepared by chloride route (left) and hydroxide route (right). Compositions correspond to (a),(b) LT; (c),(d) L3TS; (e),(f) L2S.**

As shown in Figure 2(e), chloride based samples of L2S composition were predominantly of $Li_2SiO_3$ phase with only trace signals of the orthosilicate phase observed in the form of low intensity peaks around 17°, 33.6° and 34.6° 2θ. Any coexistence of $Li_4SiO_4$ alongside the 1:1:1 $Li_2TiSiO_5$ phase cannot occur under conditions of equilibrium owing to these phases lying in separate compatibility triangles in the ternary phase equilibrium system.

For further specimen fabrication, LiCl based synthesis was ruled out owing to instability resulting from the hygroscopic nature of the chloride type material. Precipitated and dried compounds exhibit rapid decomposition and consequent phase segregation leading to deviation in stoichiometry. In similarity to other work involving chloride precursor compounds (TiCl) [27], the issue of chloride removal further discourages the use of this method for pebble fabrication.

In contrast to the chloride route, lithium loss is significantly less evident in materials synthesised using lithium hydroxide. LT samples fabricated by this method exhibit a single phase monoclinic β-$Li_2TiO_3$ composition subsequent to firing at 900°C, while samples with increasing Si content exhibit some Li deficiency evident by the presence of a metasilicate secondary phase alongside the orthosilicate target phase. It should be noted that Li excess in $Li_2TiO_3$ is a known phenomenon and thus some deviation from the projected LT-L2S tie-line is expected [29, 31, 36].

## 4.2. Spectroscopic analysis

During the synthesis of hydroxide-precursor derived material, subsequent to prolonged stirring under mild heating, sols of both L2S and LT compositions were found to exhibit a deep orange colour. This colouration was found to be stronger in L2S sols, suggesting possible formation of alkyl-lithium compounds, such as ethyllithium or methyllithium, the synthesis of which is favoured under strongly basic conditions. Consequently, ATR-FTIR spectroscopy was employed to ascertain the presence of organo-lithium compounds. Figure 2. Show FTIR scans from supernatant liquid and bulk.

Figure 3 shows FTIR spectra from sol and bulk dried precipitates of a 1:1 sol of L2S:LT. Lines indicate reported IR shifts of ethyllithium [56-58]. The large stretching vibration mode observed around 3400 cm$^{-1}$ is indicative of O-H bonds, present in solvent and precursor compounds. The lack of a significant H-O-H bending peak around 1630 cm$^{-1}$ in the sol suggests the lack of residual or structural water in the product slurry while the presence of this peak in the dried precipitates may indicate surface adsorption of water vapour. At 1098 and 1050 cm$^{-1}$ a possible C-Li stretching mode is observed. It should be noted that the formation of ethyl-lithium cannot be categorically concluded from these data, however the formation of an orange coloration may be indicative of reactions involving organolithium compounds which often degrade to form lithium hydride upon heating [59, 60].

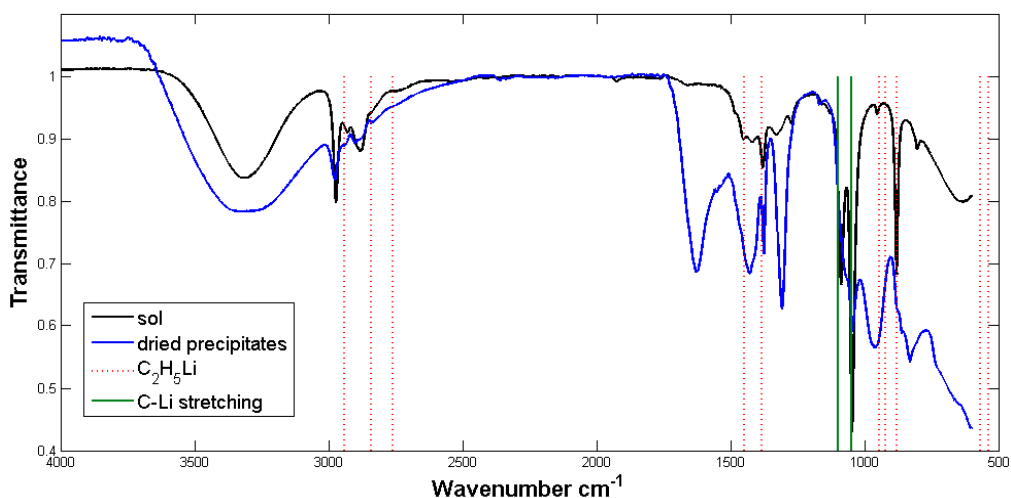

**Figure 3. FTIR spectra of combined lithium silicate and lithium titanate precursor sols (composition 7) and dried precipitates.**

XRD of powders synthesised using lithium hydroxide show the formation of lithium carbonate and Li$_2$TiO$_3$ subsequent to drying at 180° C. However, we cannot rule out the formation of lithium hydride upon drying /heating owing to the low X-ray scattering of this compound. The formation of carbonate is likely the result of a reaction with the retained organic content. A reaction with atmospheric carbon dioxide in the furnace is possible; however this is unlikely to feature prominently, particularly in materials towards the LT composition.

## 4.3. HTXRD

Temperature varied XRD (also known as High Temperature XRD, HTXRD) was employed across the range 25-1200 °C, with 50°C intervals. This was undertaken to ascertain the reaction and possible liquid formation sequence in the oxide materials. The use of eleven compositions for HTXRD analysis was carried out in order to determine the type of binary phase system exhibited over the LT-L2S range and ascertain whether any eutectic or eutectoid type behaviour is observable through the processing methods used.

For materials towards Ti rich side of the LT-L2S tie-line the dominance of Li$_2$TiO$_3$ in its monoclinic phase with space group C2/c is apparent [29, 61, 62]. This phase is found alongside lithium carbonate after drying, as shown in Figure 4, and demonstrates stability up to 1150°C with increasing crystallite size, as indicated by higher and narrower diffraction peaks. $Li_2TiO_3$ exhibits subsolidus polymorphism and above 1150°C, as predicted by reported phase equilibria studies in the $Li_2O$-$TiO_2$ system [31, 32, 36, 63], a displacive phase transformation is observed with cubic phase $\gamma$-$Li2TiO3$ solid solution, space group Fm-3m, emerging as the dominant phase at temperatures upwards of 1100°C.[51, 64] This is evident by the increase in peak intensity of the (200) peak of the rock-salt type solid solution at ~44.5° 2θ while the (002) peak of the monoclinic phase diminishes. At or below 350°C some carbonate is evident from peaks at 21.5°, 30.5° and 31.5° 2θ. At lower temperatures, during heating and cooling, the presence of small quantities of spinel phase $Li_4Ti_5O_{12}$ cannot be ruled out, owing to weak diffraction signals at 79° 2θ which can be attributed to this phase and broadening of $Li_2TiO_3$ peaks at ~35.9° and 43.6° 2θ which may indicate the overlap of (311) and (400) signals from the Ti rich phase [51]. At temperatures in the range 1000 - 1150°C multiple peaks become apparent in the range 20.5 - 25° 2θ corresponding to (111), (112), and (110) plane families in $Li_2TiO_3$ (ICDD 04-009-2812, ICSD 162215)[62]. These peaks reappear prominently during cooling and are retained to room temperature suggesting possible thermally activated preferential grain growth along these planes.

Diffraction peaks exhibit a shift to higher 2θ angles, with this trend increasing with temperature. This suggests lattice distortion which may be the result of interstitial lithium in solid solution in $Li_2TiO_3$ a phenomena which has been reported to occur up to the extent equivalent to ~44% $TiO_2$ [31].

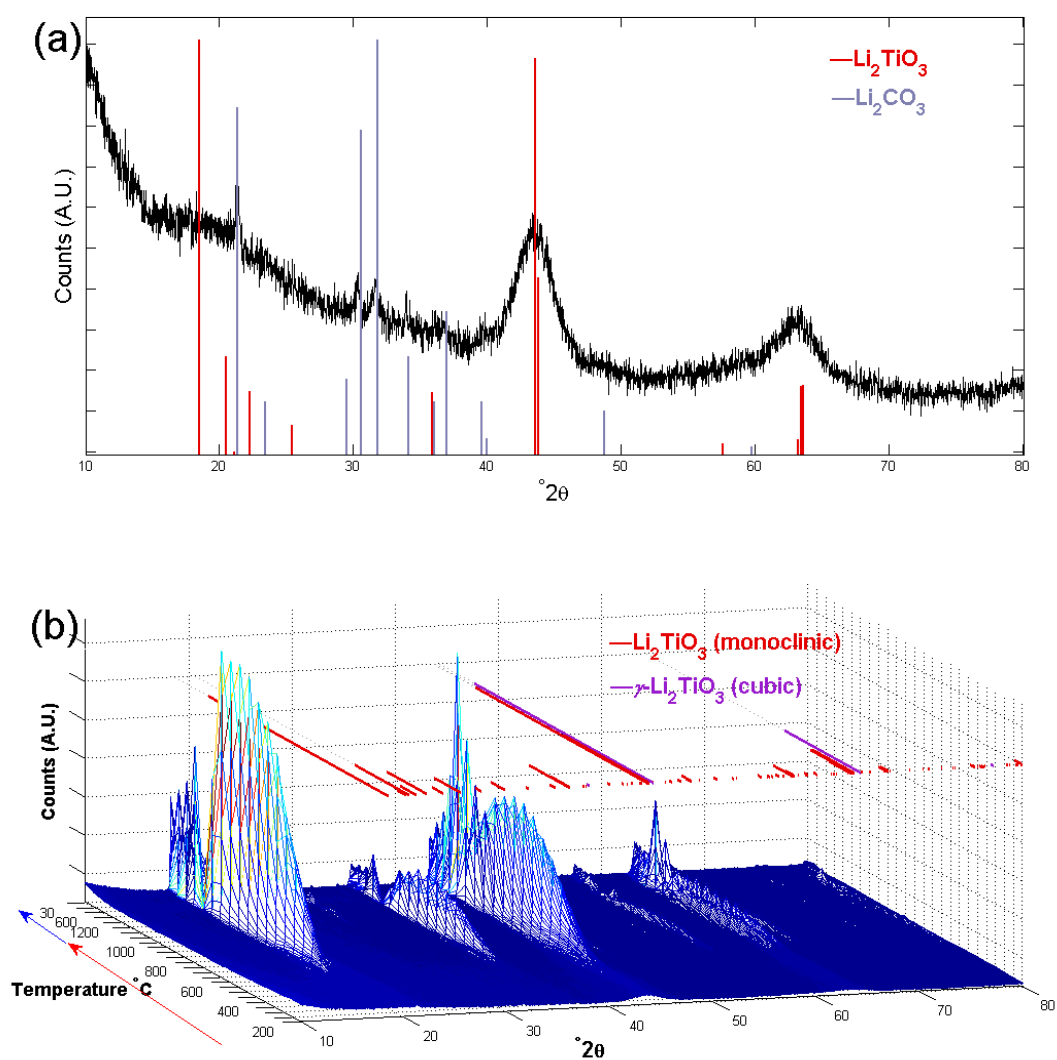

Figure 4. (a) XRD pattern for hydroxide-route-synthesised material of LT composition subsequent to drying at 180°C. (b) Temperature varied XRD for hydroxide route synthesised LT material. From front towards back shows increasing temperature, up to 1200°C, followed by cooling. Reference patterns are superimposed on horizontal plane. (Interactive figure available online)

For intermediate target compositions between LT and L2S, subsequent to drying the crystallisation of metasilicate and metatitanate phases is evident in powders, shown in Figure 5(a). With increasing lithium content diffraction peaks of carbonate material are more prominent and are present up to higher temperatures relative to compositions of lower Li content. Peaks from the orthosilicate phase are first observed in XRD patterns at 450°-500°C and this phase exhibits strong diffraction peaks in patterns obtained at temperatures at or above 550°C. At this temperature, the carbonate phase ceases to be observed in all compositions and peaks for metasilicate diminish relative to the stable metatitanate signal, suggesting a reaction between carbonate and metasilicate is involved in the formation of $Li_4SiO_4$. As temperature is increased, monoclinic $Li_2TiO_3$ peaks exhibit increasingly high intensity while peaks from the orthosilicate and metasilicate phases exhibit a lesser increase in relative intensity, suggesting a coarsening process of metatitanate crystallites. Lithium orthosilicate and metasilicate phases are not observed in diffraction patterns above temperatures of 1050°C and 1100°C respectively, while the metatitanate phase remains present. Above 1150°C a conversion to the cubic phase solid solution γ-$Li_2TiO_3$ is observed, while some peaks from the platinum crucible are evident as the result of liquid migration, as can be seen in Figure 5(b). With cooling down from 1200°C, the orthosilicate phase is not observed to recrystallise for specimens below 23.33 mole% $SiO_2$ (compositions 2-7) suggesting excess Li is retained in metatitanate and/or metasilicate solid solutions, thus impeding the recrystallization of the L2S phase. In particular The $Li_4SiO_4$ (001) and (100) peaks around 16.7° and 17.2° 2θ [65] are not prominently observed for all Ti-containing specimens subsequent to cooling, suggesting a Ti-driven modification of the growth habit of this phase.

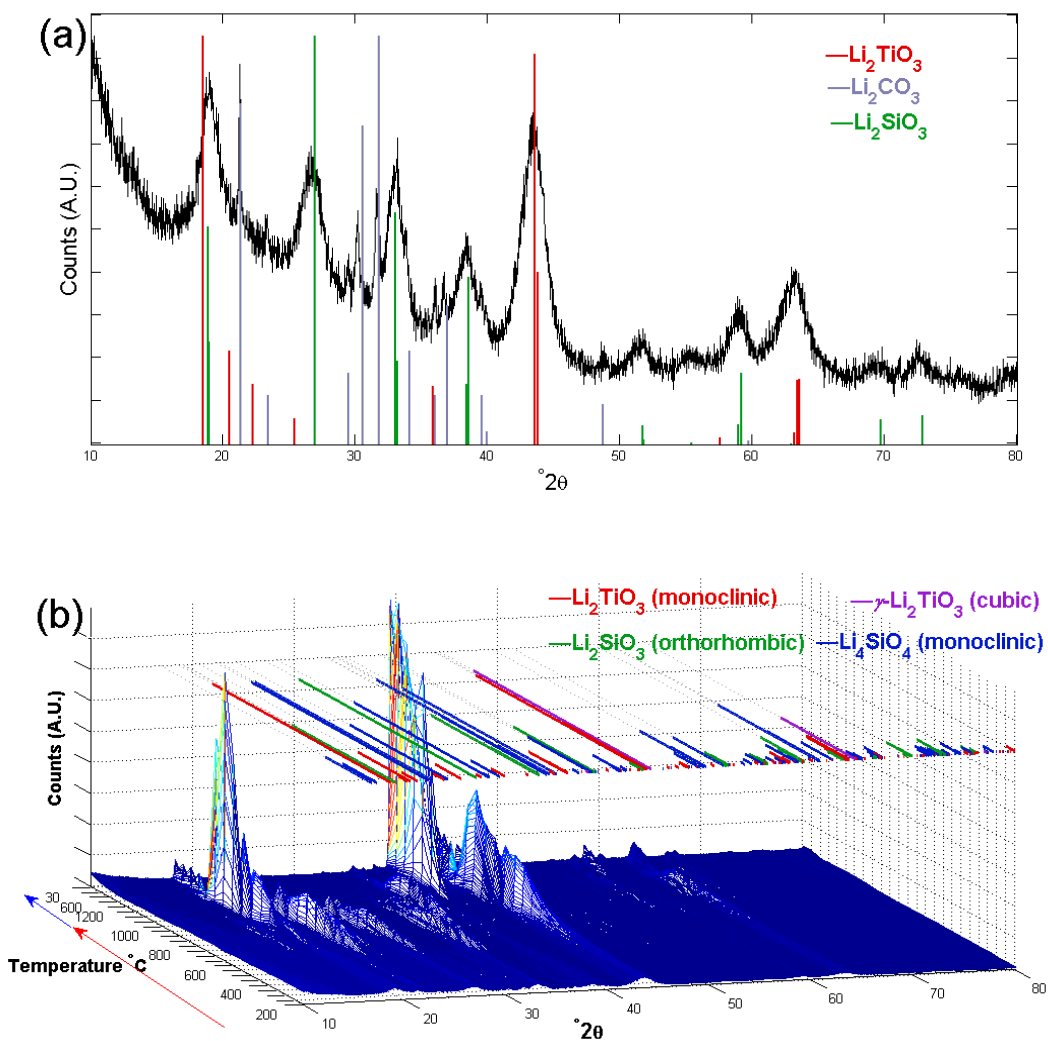

Figure 5. (a) XRD pattern for hydroxide-route-synthesised material of L3TS composition subsequent to drying at 180°C. (b) Temperature varied XRD for hydroxide route synthesised L3TS material. From front towards back shows increasing temperature, up to 1200°C, followed by cooling. Reference patterns are superimposed on horizontal plane. (Interactive figure available online)

For the sample corresponding to the L2S composition, the formation of lithium metasilicate in the as-dried material is evident by strong diffraction peaks present in as dried powder at room temperature. At temperatures below 550°C the metasilcate peaks are present alongside significant carbonate signals, as shown in Figure 6(a). This is likely owing to the unfavourable kinetics of formation of lithium orthosilicate at lower temperatures, relative to $Li_2TiO_3$ and $Li_2SiO_3$. As with intermediate compositions, a reaction between carbonate and metasilicate leads to the formation of the monoclinic orthosilicate material at temperatures between 500° and 550°C, which is the predominant phase once formed. At temperatures, between 1000° and 1050°C, $Li_2SiO_3$ peaks cease to be observed, in consistency with reported phase stability of compositions in the Li rich side of the $Li_2SiO_3$-$Li_4SiO_4$ binary system which exhibits a eutectic decomposition at 1024°C [32]. The $Li_4SiO_4$ phase exhibits strong peaks up to 1150°C. This is in contrast to Ti-bearing compositions where orthosilicate peaks diminished at lower temperatures than metasilicate. At 1200°C near complete melting of the material is evident, seen in the valley that is observed in the HTXRD stack shown in Figure 6. This occurs below the liquidus temperature (1258°C) reported in phase equilibria studies for single phase $Li_4SiO_4$ [30]. As the actual melt temperature was in the range 1150°-1200°C, this suggests an actual $Li_4SiO_4$:$Li_2SiO_3$ ratio in the region of 82:18 − 78:22 by interpolation of the liquidus surface reported in the relevant region of the $Li_2O$-$SiO_2$ phase diagram [32, 34] and equates to a lithium loss of approximately 4.5-5.5 at% relative to the target composition. As with ternary compositions, at 1200°C, subsequent to the melting of silicate phases, signals at ~39.2° and 45.7° 2θ are evident which correspond to diffraction peaks from the platinum crucible, which is exposed in the x-ray illuminated region owing to significant melt formation and its migration away from the crucible centre. In contrast to intermediate compositions, upon cooling of composition 11 from 1200°C down to room temperature, the orthosilicate phase (corresponding to the target L2S ratio) recrystallises while the metasilicate phase remains notably less prominent, potentially owing to the formation of amorphous silicate phases. Peaks from the exposed platinum crucible remain during cooling.

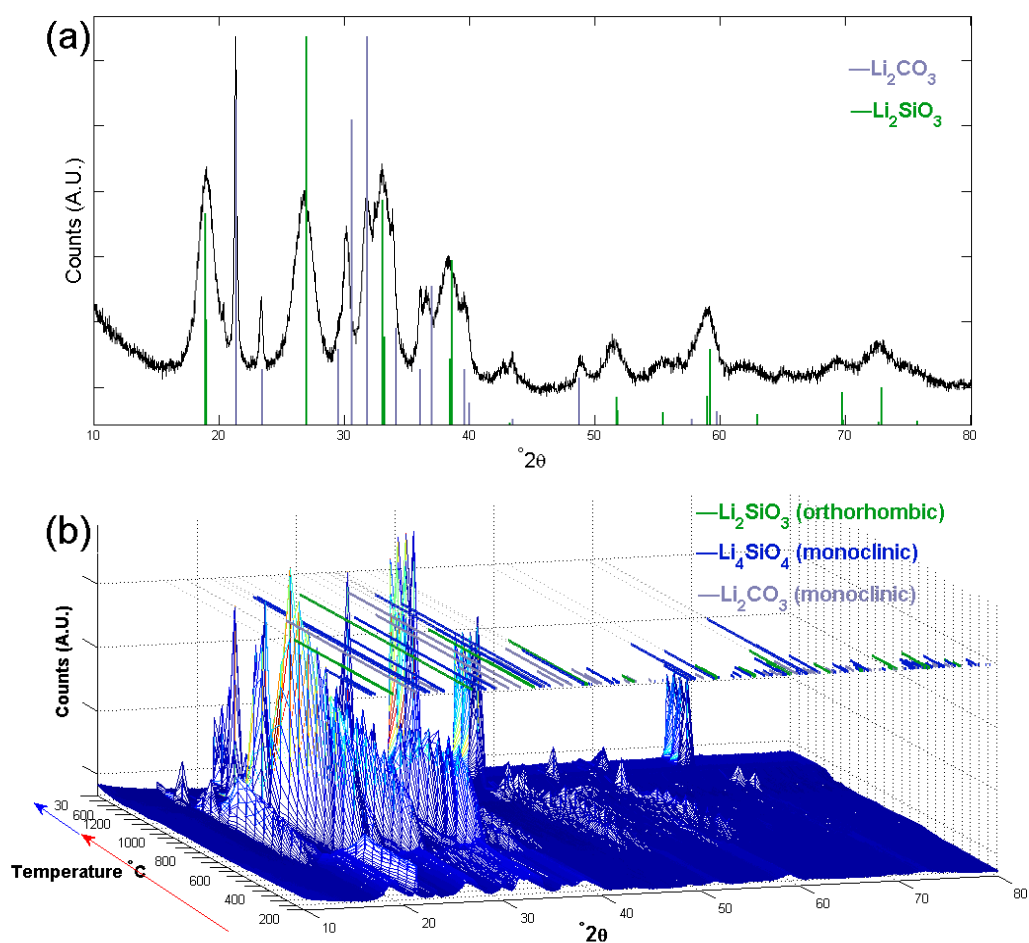

Figure 6. (a) XRD pattern of L2S material fabricated by hydroxide route subsequent to drying at 180°C (b) Temperature varied XRD for hydroxide route synthesised LT material. From front towards back shows increasing temperature, up to 1200°C, followed by cooling. Reference patterns are superimposed on horizontal plane. (Interactive figure available online)

The lack of observed melting of $Li_2TiO_3$ in compositions towards the orthosilicate side of the studied system suggests that the LT-L2S quasi-binary is a monotectic rather than a eutectic system, this is further shown in the behavioural diagram in Figure 8.

### 4.4. Thermogravimetry / DSC

Thermogravimetry (TG) and differential scanning calorimetry (DSC) were carried out in order to further examine the temperature and composition dependant transitions observed by XRD. Results for the endpoint and median compositions 1, 7 and 11 (LT, L3TS and L2S) are shown in Figure 7. These data confirm a significant loss of mass during heating up to 800°C, evident in the mass loss of approximately 17%, 19% and 26% in powders of LT, L3TS and L2S compositions respectively, as the result of the pyrolysis of residual organic content and $Li_2CO_3$ in this temperature range. In L2S specimens a further ~2.5% mass loss occurs in the range 1150°-1350 °C and this is attributed to the near-complete melting observed in this composition and the accompanying further Li volatilisation.

Calorimetric data, plotted at different scales, show exothermic peaks in the region 300°-400°C with these peaks being more significant in titanate bearing compositions. The greater exothermic behaviour in compositions richer in Ti is likely the result of precursor chemistry leading to a greater organic content in these powders. It is worth noting that in $Li_2TiO_3$ exothermic behaviour at ~400°C is also likely to be influenced by the transition from α-$Li_2TiO_3$ to β-$Li_2TiO_3$ which occurs around this temperature range [66]. The formation of a metastable cubic α-$Li_2TiO_3$ at low temperatures which subsequently transforms to the stable monoclinic phase at 350°-400°C was recently reported by Laumann et al. in a series of publications [67-69], however this phase transformation is not readily observable in XRD scans as α-$Li_2TiO_3$ diffraction peaks coincide with signals from the monoclinic β-phase [68].

Regions of interest are numbered on the DSC scan data shown in figure 7. **(1)** Endothermic removal at 100°-200°C of physisorbed moisture **(2)** Exothermic combustion at ~200°-400°C of residual organic groups (propoxide and ethoxide from TTIP and TEOS respectively) and α to β $Li_2TiO_3$ phase transformation. **(3)** Endothermic reaction at composition dependant temperatures in the range 520°-580°C of $Li_2CO_3$ to form $Li_2TiO_3$ and, more significantly, $Li_2SiO_3$ accompanied by mass loss through release of $CO_2$ [70] **(4)** Endothermic behaviour with fluxed melting of residual $Li_2CO_3$ (<724°C) accompanied by further silicate/titanate formation following reported reactions [70, 71]. **(5)** Endothermic behaviour at temperatures above 900°C resulting from grain growth, phase transformation in $Li_2TiO_3$, and melting of orthosilicate. **(6)** Strongly exothermic crystallisation of $Li_2TiO_3$, $Li_2SiO_3$ and $Li_4SiO_4$ during cooling from 1350°C down to 600°C as well as the γ to β-$Li_2TiO_3$ phase transformation **(7)** Exothermic behaviour accompanied by slight mass increase around 400°C during cooling of silicate containing compositions, potentially resulting from the reaction of $Li_4SiO_4$ with atmospheric $CO_2$, which is reported to occur most rapidly around this temperature regime [72].

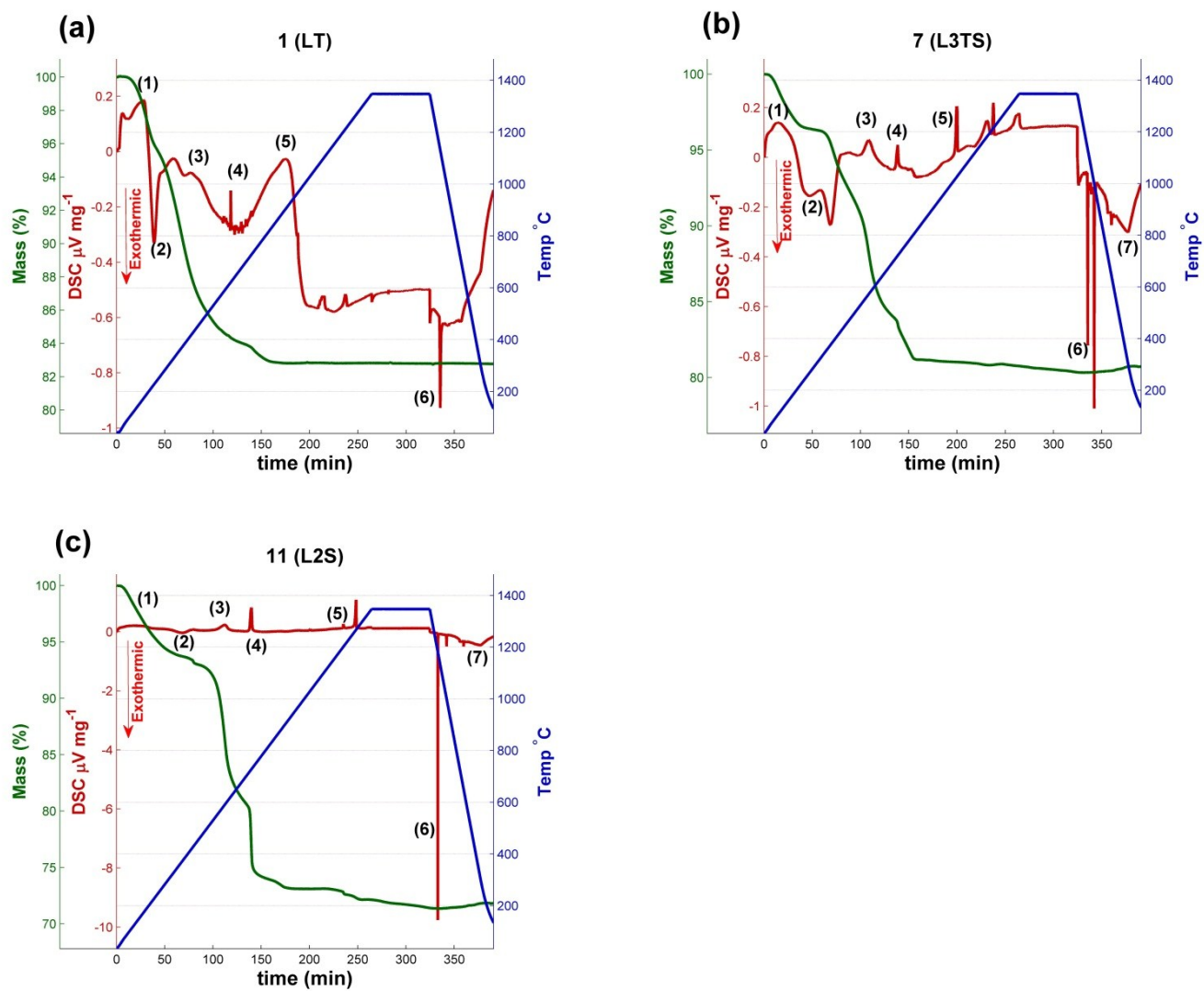

**Figure 7. TG/DSC profiles for (a) LT material, (b) L3TS material and (c) L2S material.**

*4.5.    Summary of phase transformation behaviour*

The behavioural diagram shown in Figure 8 presents a summary of observed phase transformation behaviour during heating, compiled from the combination of XRD and DSC data for materials fabricated from hydroxide sol precursors. As temperatures above 1350°C were not investigated, higher temperature behaviour shown is an interpretive projection. The shown transformation of α- to β-$Li_2TiO_3$ at 400°C is somewhat speculative as XRD results obtained in the present work cannot confirm or dispel the existence of this phase in synthesised materials reported.

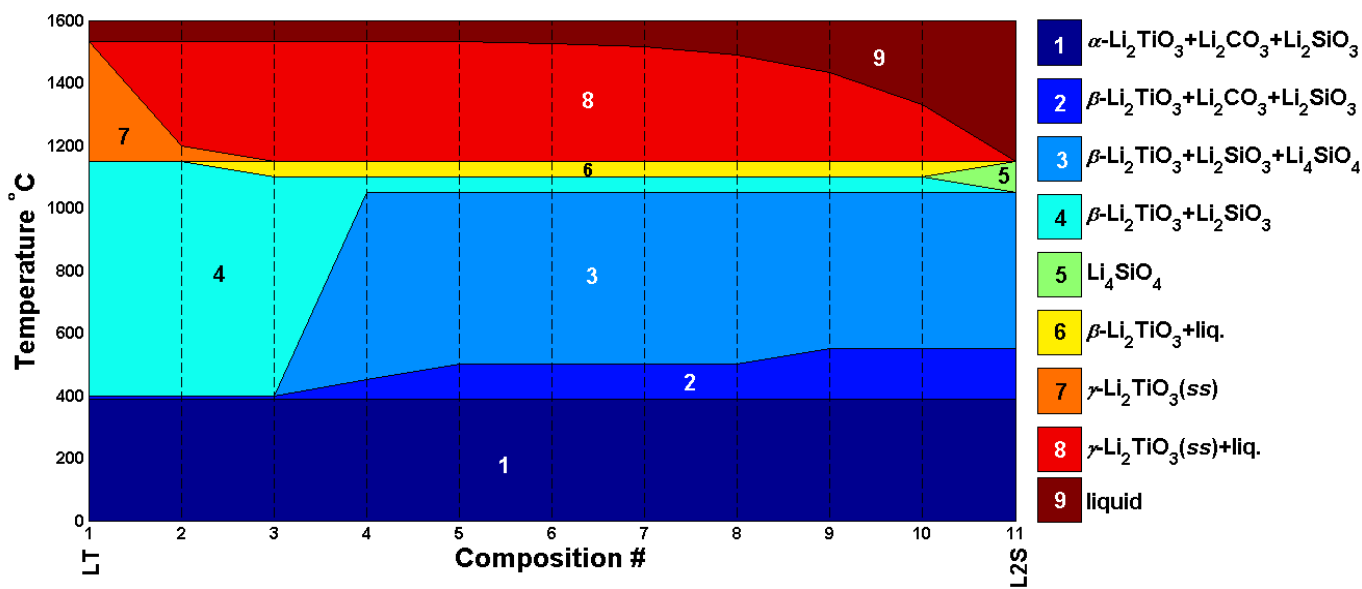

**Figure 8. Behavioural phase transformation diagram from a summary of temperature varied XRD and DSC data of the 11 compositions studied in heating mode.**

## 5. Discussion

### 5.1. Synthesis route and Li deficiency

The use of chloride precursors is frequently reported in sol-gel processing of metal oxides and allows better pH control in sols relative to hydroxide precursors. In the present work the use of this compound was found to be unfavourable owing to a resulting stoichiometric imbalance. Additionally the use of chloride is problematic for HCPB materials as residual chloride is undesirable in breeder blankets [73].

The greater extent of Li loss in the chloride based materials may result from the hygroscopic nature of the precursor and consequent segregation of lithium towards the material surface during drying. This lithium deficiency manifested in the crystallisation of Li poor phases including $Li_2TiSiO_5$ and $Li_4Ti_5O_{12}$ spinel. Alternative soluble Li precursors include $LiNO_3$, $Li_2SO_4$ and $LiC_2H_3O_2$, while alternative Si and Ti precursors too are available and, subject to further investigation, may be employed to minimise Li loss to allow control of stoichiometry, morphology and crystallinity. Additionally processing procedures and parameters can be varied, facilitating the fabrication of compositions and microstructure different to those obtained in the present work.

In hydroxide synthesised materials, the presence of metasilicate alongside the target orthosilicate phase is a possible result of two main causes, the incorporation of excess lithium into $Li_2TiO_3$, as reported elsewhere, and/or the sublimation of lithium (as atomic lithium or gaseous lithia) from $Li_4SiO_4$ leading to the formation of $Li_2SiO_3$ as reported by Cruz et. al. [45]

The phenomenon of lithium loss through volatilisation in analogous systems is well known [4, 44]. The presence of lithium in glassy phases cannot be ruled out, however reported Li-Si glass systems are significantly richer in Si than the phases investigated in the present work, and hence their formation would not account for the lack of orthosilicate phase in compositions studied here relative to the target compositions [74, 75], moreover one would expect rapid devitrification of these systems to result from the applied thermal treatments [76]. In materials fabricated using the present methods with a hydroxide precursor, from analysis of liquid formation temperature, at which the metasilicate phase is observed to melt, it can be inferred that the loss of lithium in the L2S composition was in the region of ~5% This implies that the use of excess lithium as employed by Hoshino et al. [1, 4, 77] can be used to correct for this occurrence and facilitate the attainment of biphasic metatitanate-orthosilicate mixtures. It should be noted that the precise loss of lithium and the excess lithium necessary to correct for this is dependent on specimen parameters and firing conditions.

### 5.2. Precursor sol-chemistry

The appearance of an aromatic orange compound in the LiOH based sols suggests a possible formation of organo-lithium compounds which merits further investigation. The decomposition of such compounds during drying is likely, resulting in possible hydride or carbonate formation.

### 5.3. Crystallisation

The formation of metasilicate and metatitanate phases occurs in nano-crystalline form subsequent to drying, as indicated by the wide diffraction peaks evident for these phases at low temperatures. The observed co-formation of carbonate can likely be avoided by use of alternative sol-gel processing using aqueous sols or otherwise removing organic content. While nanostructured material can be advantageous for tritium release characteristics, the expected application temperature in breeder blankets is in the region of 900°C, and thus nano-crystallinity is unlikely to be preserved in the absence of grain-growth limiting dopants, the use of which would be undesirable in breeder pebbles. DSC data indicates that the hydroxide route employed in the present work may result in the crystallisation of metastable cubic $\alpha$-$Li_2TiO_3$ in similarity to results obtained by hydrothermal processing by Laumann et. al. [67, 68]. Ascertaining the formation of the metastable phase is attainable by synchrotron or neutron diffraction studies, however for breeder-pebble applications this phase is of little significance.

### 5.4. Phase stability

Observed compositions of $Li_2TiO_3$ existing alongside a phase mixture of $Li_4SiO_4$ and $Li_2SiO_3$ are similar to 3-phase compositions obtained by melt based breeder pebble processing [13] and are consistent with kinetic studies in the Li metasilicate-orthosilicate regime [45] and with studies into synthesis methods of lithium silicate powders [28, 33]. The coexistence of lithium metatitanate with predominantly orthosilicate material infers compatibility between L2S and LT regions of the ternary system. The solidus temperature in this system appears to be in the range 1050°-1100°C while further studies are required to describe fully the liquidus surface in this regime. Based on the range of 11 compositions evaluated, the quasi binary system is likely to be of a monotectic type as no evidence of $Li_2TiO_3$ melt existing alongside solid $Li_4SiO_4$ was observed at any composition.

The absence of orthosilicate in compositions 2 and 3 during the heating stage of HTXRD analysis indicates a likely solid solution of excess lithium in the $\beta$-metatitanate phase. Such non-stoichiometry towards both Li and Ti rich compositions has frequently been reported in studies in the $Li_2O$-$TiO_2$ system [31, 36]. Additionally the recrystallization of orthosilicate was not clearly observed upon cooling from a composition of $\gamma$-$Li_2TiO_3$+liquid down to room temperature for intermediate compositions (samples 2-7). This may further indicate solid solution with excess Li incorporated into $Li_2TiO_3$ and/or $Li_2SiO_3$ although the evaporation of Li from the liquid phase is an additional contributing factor.

## 6. Conclusions

Biphasic mixtures in the quasi-binary $Li_2TiO_3$-$Li_4SiO_4$ system offer the scope for improved properties, relative to single phase materials, for tritium breeding applications in fusion reactors. For this reason the solution based synthesis of materials in this system was studied using organometallic precursors with a focus on crystallisation and phase stability behaviour. Owing to deviation from target stoichiometries, the presently reported procedures did not result in purely biphasic $Li_4SiO_4$-$Li_2TiO_3$ mixtures however it is likely that such two-phase compositions are attainable through solution based methods employing appropriate processing parameters including the addition of excess lithium. In addition, alternative fabrication methods warrant future investigation and in particular the use of inorganic precursor compounds merits further research in order to shed light on the lithium rich region of the ternary $Li_2O$-$SiO_2$-$TiO_2$ system, which remains poorly understood. In the synthesis procedures reported here we have identified several issues which require consideration for future solution-based syntheses of mixed phase materials.

- Lithium sublimation and its dependence on firing conditions and composition
- Non-stoichimetry in $Li_2TiO_3$, particularly at high temperatures
- Ubiquitous formation of metasilicate owing to favourable crystallisation kinetics of this phase
- Potential formation of organolithium products in precursor sols, which may be controlled through parameters of time, temperature, pH and composition
- Crystallisation of carbonate through reaction with retained organic content
- The existence of a likely monotectic system with liquid formation at ~1050°C.

**Acknowledgements**

This work was supported by the Group of Eight - DAAD Joint Research Cooperation Scheme.